\documentstyle[prl,aps,multicol,epsfig]{revtex}

\begin{document}

\def\ra{\rangle}
\def\la{\langle}
\def\bege{\begin{equation}}
\def\ende{\end{equation}}
\def\begarr{\begin{eqnarray}}
\def\endarr{\end{eqnarray}}
\def\no{\noindent}\def\non{\nonumber}
\def\h{{1\over 2}}
\def\hi{\hangindent=45pt}
\def\v{\vskip 12pt}

\draft

\title{
From Laser Induced Line Narrowing
To 
Electromagnetically Induced Transparency:\\
Closed System Analysis
}

\author{%
Hwang~Lee,$^{1}$\footnote{
Present address:
Jet Propulsion Laboratory, MS 126-347,\\
California Institute of Technology,
Pasadena, CA~91109
}
Yuri~Rostovtsev,$^{1}$
Chris~J.~Bednar,$^{1,2}$
and Ali~Javan$^{2,3}$
}

\address{
$^1$ Department of Physics, Texas A\&M University,
        College Station, TX~~77843 \\
$^2$ Max-Planck-Institut f\"{u}r Quantenoptik,
        D-85748 Garching, Germany \\
$^3$Department of Physics,
      Massachusetts Institute of Technology,
      Cambridge, MA~~02139
}

\date{February 1, 2002}

\maketitle

\begin{multicols}{2}


{\bf Abstract.}
Laser induced line narrowing effect,
discovered more than thirty years ago,
can also be applied to recent studies in high resolution
spectroscopy based on electromagnetically induced
transparency.
In this paper we first present a general form of the 
transmission width of electromagnetically induced
transparency in a homogeneously broadened medium.
We then analyze a Doppler broadened medium
by using a Lorentzian function as the atomic velocity
distribution.
The dependence of the transmission linewidth 
on the driving field intensity
is discussed and compared to
the laser induced line narrowing effect.
This dependence
can be characterized by
a parameter
which can be regarded as ``the
degree of optical pumping''.

~

\no

~

\hrulefill

~

Over the last decade, 
considerable attention
has been paid to the studies of
the atomic coherence effects
and their applications \cite{arimondo96,harris97}.
The technique of Electromagnetically Induced Transparency (EIT) 
which makes an opaque medium become transparent
by applying an external coherent radiation 
field \cite{olga-eit,boller91},
yields
various applications from
enhancement of nonlinear
optical processes \cite{harris90,hakuta91,harris99},
to slow light
\cite{harris92,xiao95,schmidts96,hau99,kash99,budker99,olga01}.
In addition to the elimination of absorption,
the absorption profile
reveals
a narrow transmission line, which
has been applied to 
high resolution spectroscopy and 
high sensitivity magnetometer 
\cite{mos92,brandt97,lukin97,budker98}.

Since many of these experiments are performed
in an atomic cell configuration,
the Doppler broadening effect on EIT
is an important concern.
Recent theoretical investigations of
Doppler broadening effects on EIT, however,
has been focused mainly on
the existence of EIT for certain 
configurations \cite{julio95,kara95,wang95}.
The issue of EIT linewidth for a Doppler 
broadened medium
has been lately addressed by Taichenachev
and coworkers \cite{taich00}.
As the width of transmission line
is directly related to the dispersion
near the EIT resonance,
it is also a key issue in
dispersive measurements.

In a three-level $\Lambda$-type system
if the system is homogeneously broadened,
as is well known,
EIT can be achieved when the intensity 
of the driving field ($\Omega^2$) is larger than
the product of
the decay rate of the
coherence between the lower levels ($\gamma_{bc}$)
and the homogeneous linewidth ($\gamma$).
Then, 
if the system is inhomogeneously broadened
(say, with the width $W_D$),
one might guess that
EIT can be achieved 
when $\Omega^2$ is larger than
$\gamma_{bc} W_D$ instead of $\gamma_{bc}\gamma$.
This is not so.
We show that one can still have EIT when
$\Omega^2 \gg {\gamma_{bc}\gamma}$
even in the case of inhomogeneous broadening.

For the spectral width of EIT,
if the system is homogeneously broadened,
the two absorption lines are
separated approximately by the
Rabi frequency of the driving field $\Omega$ 
when $\Omega$ is larger than the homogeneous linewidth $\gamma$.
When $\Omega \ll \gamma$,
it becomes $\Omega^2/\gamma$.
Then, if the system is inhomogeneously broadened,
it might be inferred that the EIT width
goes as $\Omega$ when $\Omega$ is larger than
the inhomogeneous linewidth $W_D$,
and becomes $\Omega^2/W_D$
as $\Omega \ll W_D$.

In the literature, however, we find that
the narrow feature superimposed on
the Doppler broadened profile
has been studied more than thirty years ago.
Laser induced line narrowing effect
was discovered by Feld and Javan \cite{feld69}
and the spectral width of
the narrow line was shown to be linearly
proportional
to the driving field Rabi frequency.
Various aspects of this effect
has been investigated
by H\"ansch and Toschek \cite{hansch70}, and
it was also called nonlinear interference effects
\cite{popova70}.
In a recent article \cite{javan00},
it has been
proposed that 
this laser induced line narrowing can be 
applied to the recent experiments
based on EIT and
the spectral line of
the EIT resonance
can be narrower in a Doppler broadened system
than in a homogeneously broadened system.
Here we 
analyze these ideas in detail
and demonstrate 
the power broadening of
the linewidth of EIT resonance
in a Doppler broadened system.

Under the condition of $\Omega \ll W_D$,
there are again two different regimes
of EIT width:
In one limit
it is proportional to the Rabi frequency
of the driving field,
which has the same expression as the spectral
width shown in the study of
laser induced line narrowing \cite{feld69}.
As the driving field gets strong,
it becomes power broadened
and indeed has a form proportional to the intensity of
driving field (as $\Omega^2/W_D$).

This paper is organized as follows:
In Sec.~I we set up our model scheme of
the three-level system 
and the transmission width of EIT in a homogeneously 
broadened medium is discussed.
In Sec.~II the Doppler averaged susceptibility is
obtained by using a Lorentzian function for the velocity distribution
and the absorption profile, the EIT condition, and the linewidth
of EIT are discussed.
Comparison between the closed system and the open system 
is briefly given in Sec.~III.
Section IV contains the summary of
the present paper.

\section{Homogeneously Broadened System}

We consider a model scheme depicted in Fig.~1.
The transition $a \leftrightarrow c$ is coupled to a
coherent driving field and 
the transition $a \leftrightarrow b$ is coupled to a
weak probe field.
The atom-field interaction Hamiltonian
can be written as

\bege
{\cal V}=-\hbar \alpha e^{-i\nu t} |a \ra \la b|
-\hbar \Omega e^{-i\nu_0 t} |a \ra \la c| + {\rm H.c.}  ,
\label{h1}
\ende

\no
where $\alpha$ is the Rabi frequency of the probe field,
$\Omega$ are the Rabi frequency of the driving field.
In this model we take the decay rate from the level $a$ to $b$ ($c$) as
$\gamma$ ($\gamma^\prime$).
The relaxation between the lower levels
is denoted as $\gamma_{bc}$
such that 
the decay rate of the off-diagonal density matrix element
($\rho_{bc}$) is defined as $\gamma_{bc}$.

\begin{figure}[htb]
\epsfysize=3.5cm
\centerline{\epsfbox{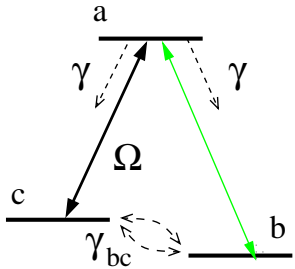}}~
\caption{\label{fig-1}
Three-level model scheme.
The upper level $a$ decays to $b$ and
$c$ with decay rate $\gamma$.
The relaxation rate between levels $b$ and $c$
is denoted as $\gamma_{bc}$ which is assumed to be small
compared with $\gamma$.
}
\end{figure}

The equations of motion for the density matrix elements
in a rotating frame are then given by

\begin{mathletters}
\label{h2}
\begarr
{\dot\rho}_{ab}&=& -  {\Gamma}_{ab}\rho_{ab}
           -i \alpha (\rho_{aa}-\rho_{bb}) +i \Omega \rho_{cb}
\label{a} \\
{\dot\rho}_{cb}&=&  -  {\Gamma}_{cb}\rho_{cb}
           -i \alpha \rho_{ca} +i \Omega \rho_{ab}
\label{b} \\
{\dot\rho}_{ac}&=&  -  {\Gamma}_{ac}\rho_{ac}
           -i \alpha \rho_{bc} -i \Omega (\rho_{aa}-\rho_{cc})
\label{c} \\
{\dot\rho}_{cc}&=&  -\gamma_{bc} \rho_{cc}
           + \gamma' \rho_{aa} + \gamma_{bc} \rho_{bb} 
           - i\Omega (\rho_{ca}- \rho_{ac})
\label{d} \\
{\dot\rho}_{aa}&=& -  (\gamma+\gamma') \rho_{aa}
           - i\alpha( \rho_{ab} -\rho_{ba} ) 
           - i\Omega (\rho_{ac}- \rho_{ca}) .
\label{e}
\endarr
\end{mathletters}

\no
Here we
assume that the Rabi frequencies are real,
$\Gamma_{ij}$'s are defined as $\gamma_{ij} + i \Delta_{ij}$,
where

\begarr
\gamma_{ab}  &=& \gamma_{ac} =
        \h(\gamma + \gamma' + \gamma_{bc}) ,
\qquad
\gamma_{cb}  = \gamma_{bc} .
\label{h3}
\endarr

\no
and $\Delta_{ij}$'s are given as
$\Delta_{ab}=\omega_{ab}-\nu$,
$\Delta_{ac}=\omega_{ac}-\nu_0$,
and $\Delta_{cb}=\Delta_{ab}-\Delta_{ac}$.

For a weak probe field,
first order solution for the off-diagonal density 
matrix element $\rho_{ab}$ 
(which governs the absorption of the probe field)
can be found
in steady state as

\bege
\rho_{ab}^{(1)}={-i\alpha \over \Gamma_{ab}\Gamma_{cb} +\Omega^2}
\Big[ \Gamma_{cb} (\rho_{aa}^{(0)}-\rho_{bb}^{(0)})
+ {\Omega^2 \over \Gamma_{ca}}
(\rho_{cc}^{(0)}-\rho_{aa}^{(0)}) \Big] .
\label{h6}
\ende

\no
where
$\rho_{ll}^{(0)}$ is the population in level $l$
in the absence of the probe field.
The susceptibility is then written as

 \begin{equation}
  \chi
     =
\eta
        \left\{ {\rho_{ab}^{(1)} \over \alpha} \right\},
  \label{chi-1}
 \end{equation}
where
$\eta$ is given by
${\eta \equiv (3/8\pi) N \gamma \lambda^3}$  
for the atomic number density $N$
and the wavelength $\lambda$.
The effect of the probe field intnsity
on the susceptibility 
is ignored by using the linear approximation
\cite{taich00}.

\subsection{Optical pumping and population distribution}

Let us find
the population of each level
in the absence of the probe field (i.e. the
zeroth order population).
Obviously, if the driving field is not turned on,
we have $\rho_{aa}=0$
and $\rho_{bb}=\rho_{cc}=1/2$
from Eq.~(\ref{h2}).
Now as the driving field being turned on,
in steady state, we have
from Eq.~(\ref{h2}c,e)

\begarr
\rho_{ac}^{(0)} &=& -{i\Omega \over \Gamma_{ac} }
(\rho_{aa}^{(0)} - \rho_{cc}^{(0)}) , \non \\
\rho_{aa}^{(0)} &=&  -{i\Omega \over \gamma + \gamma' }
(\rho_{ac}^{(0)} - \rho_{ca}^{(0)}) .
\label{ss}
\endarr

\no
Let us now assume, for the sake of simplicity,
that the decay rate from the level $a$ to $c$
is same as the decay rate from the level $a$ to $b$,
i,e.~$\gamma' = \gamma$ and the driving field detuning
is denoted as $\Delta_0$.
Then, we have
$\Gamma_{ac} = \gamma_{ac} +i \Delta_{ac}
= (2 \gamma + \gamma_{bc})/2 + i \Delta_0$,
and

\bege
\rho_{ac}^{(0)} - \rho_{ca}^{(0)}
= {- i 2 \Omega (\gamma + \gamma_{bc}/2) \over 
(\gamma +\gamma_{bc}/2)^2 + \Delta_0^2 }
\left(\rho_{aa}^{(0)} - \rho_{cc}^{(0)}\right)  .
\label{rhoac}
\ende

\no
By Eqs.~(\ref{ss}) and (\ref{rhoac})
we obtain

\bege
\bigg[ 2 \gamma  + {\Omega^2 \over X} \bigg]
\rho_{aa}^{(0)} =
{\Omega^2  \over X} \rho_{cc}^{(0)}  ,
\ende

\no
where
\bege
  X \equiv
  {
   [(\gamma + \gamma_{bc}/2)^2 + \Delta_0^2] \over
   2 (\gamma +\gamma_{bc}/2) }.
\label{x}
\ende

\no
Note
that
Eq.~(\ref{h2}d) can be written as

\begarr
{\dot\rho}_{cc}
 &=&  - \left( \gamma_{bc} + {\Omega^2 \over X} \right) \rho_{cc}
      + \left(\gamma  + {\Omega^2 \over X} \right) \rho_{aa} 
      + \gamma_{bc} \rho_{bb}
\label{op}
\endarr

\no
Hence using $\rho_{bb}=1-\rho_{aa} -\rho_{cc}$,
we obtain the zeroth order population


\begin{mathletters}
\label{h4}
\begin{eqnarray}
\rho_{aa}^{(0)} &=& { 2 \gamma_{bc} \Omega^2 \over 2 D} ,
\\
\rho_{bb}^{(0)} &=& { 4 \gamma X \gamma_{bc} + 2 \gamma_{bc} \Omega^2
 + 2 \Omega^2 \gamma \over 2D } ,
\label{h4a} \\
\rho_{cc}^{(0)} &=& { 4 \gamma X \gamma_{bc} + 2 \gamma_{bc} \Omega^2
  \over 2D },
\label{h4c}
\end{eqnarray}
\end{mathletters}
where
%
$
  D
    \equiv
        4 \gamma_{bc} \gamma X + 3 \gamma_{bc} \Omega^2
         + \Omega^2 \gamma  .
$
%
For $\gamma \gg \gamma_{bc}$,
these can be simplified as

\begin{eqnarray}
\rho_{aa}^{(0)} -\rho_{cc}^{(0)} 
 &\approx& - { 4 \gamma X \gamma_{bc} \over 2D },
\quad
\rho_{bb}^{(0)} 
\approx { 4 \gamma X \gamma_{bc} + 2 \Omega^2 \gamma \over 2D } ,
\label{pop}
\end{eqnarray}

\no
where

\begarr
X &\approx& {\gamma^2 + \Delta_0^2 \over 2 \gamma},
\qquad
D \approx
        4 \gamma_{bc} \gamma X + \Omega^2 \gamma  .
\label{xx}
\endarr

\no
Note that when the driving field is on resonance,
the usual EIT condition
$\Omega^2 \gg \gamma_{bc} \gamma$
is equivalent
to
$\rho_{bb}\approx 1$
in Eq.~(\ref{pop});
i.e.~a complete
optical pumping to
the level $b$ is
required to achieve
EIT.

\subsection{Transmission width of EIT}

Now let us consider the transmission width
under the condition of a resonant driving field.
When we have a resonant
driving field, i.e.~$\Delta_0=0$,
from Eq.~(\ref{xx}).
we find $X \approx \gamma/2$ and
$D \approx \Omega^2 \gamma$.
Therefore,
$\rho_{aa}^{(0)} \approx
\rho_{cc}^{(0)} \approx 0$
and $\rho_{bb}^{(0)} \approx 1$, i.e.~all
the populations are in the level $b$.
As is discussed in the previous section,
the condition $\Omega^2 \gg \gamma_{bc} \gamma$
leads to a complete optical pumping
in the homogeneously broadened case.

Eqs.~(\ref{h6},\ref{chi-1}) then yield

\begarr
\chi & = &
{ \eta (-i)\Gamma_{cb} (-1) 
\over \Gamma_{ab} \Gamma_{cb} +\Omega^2}  .
\label{b1}
\endarr

\no
Since $\Gamma_{ab} \approx \gamma +i \Delta$
and $\Gamma_{cb} = \gamma_{bc} +i \Delta$,
we have

\begarr
\chi & = &
{\eta i\over Z}
(\gamma_{bc} + i \Delta) 
\bigg[ (\Omega^2 -\Delta^2) - i\Delta \gamma \bigg] ,
\label{b2}
\endarr

\no
where $Z=(\Omega^2 -\Delta^2) + \Delta^2 \gamma^2$.
Hence,
the imaginary part is obtained as

\begarr
\chi'' & = &
{\eta \over Z}
\bigg[ \gamma_{bc} (\Omega^2 -\Delta^2) +
\Delta^2 \gamma \bigg] .
\label{b3}
\endarr

\no
Since the maximum of $\chi''$ is
$1/\gamma$ at $\Delta \approx \Omega$,
we may define $\Gamma_{EIT}$, the half width
of EIT as
$\chi''(\Delta =\Gamma_{EIT}) = 1/2\gamma$,
which gives

\bege
\Delta^4
- \Delta^2 (2 \Omega^2 + \gamma^2)
+ \Omega^4 =0 ,
\ende

\no
and the solution is

\bege
\Delta^2 =
{\gamma^2 \over 2}
\bigg[
2 s + 1 \pm \sqrt{4 s + 1} \bigg] ,
\ende

\no
where
$s = \Omega^2 /\gamma^2$.
Hence for $s \gg 1$ we have

\begarr
\Delta^2 &\approx& \gamma^2 (s + \sqrt{s})
\non \\
\Rightarrow
\Delta &\approx& \pm~\Omega \pm {\gamma \over 2} ,
\endarr

\no
which shows that the absorption peaks are at
$\pm \Omega$ with full width $\gamma$
and the half width of transmission is
obtained as

\begarr
\Gamma_{EIT} &\approx& \Omega - {\gamma \over 2} .
\endarr

On the other hand,
for $s \ll 1$
we have

\begarr
\Delta^2 &\approx&
{\gamma^2 \over 2}
\bigg[
2 s + 1 \pm \left( 1 + 2 s -2 s^2 \right) \bigg] .
\endarr
\no
Therefore,
\begarr
\Longrightarrow~~~~~~
\Delta &\approx&
\pm \left( \gamma + {\Omega^2 \over \gamma} \right),
\qquad
\pm {\Omega^2 \over \gamma} 
\endarr

\no
Hence, when $\Omega \ll \gamma$,
we have the absorption profile showing
a whole envelope with half width
$\gamma + \Omega^2/ \gamma$, and at the center
there exists a transmission line
with its half width as

\bege
\Gamma_{EIT} \sim \Omega^2/\gamma.
\ende

\no
We note that under the EIT condition 
$\Omega^2 \gg \gamma \gamma_{bc}$,
$\Gamma_{EIT}$ cannot be smaller than
$\gamma_{bc}$.

\section{Inhomogeneouly Broadened System}

Now if the system is Doppler broadened,
for the atoms with velocity $v$,
the radiation fields are Doppler shifted
as $\nu \rightarrow \nu (1 - v/c)
= \nu - k v$ for the probe field
with $k$ as the component of the wavevector
on the propagation axis, and
$\nu_0 \rightarrow \nu_0 (1 - v/c)
= \nu_0 - k' v$ for the driving field.
Hence, for a Doppler broadened
system, we replace
$\Delta_{ij}$ as
$\Delta_{ab} \rightarrow \Delta_{ab} + k v$
$\Delta_{ac} \rightarrow \Delta_{ac} + k' v$,
and
$\Delta_{cb} \rightarrow \Delta_{cb} + (k -k') v$.
In the present analysis we assume that
the energy difference between the level $b$ and $c$
is small enough so that we have $k' \approx k$
and the probe field and the driving
field are copropagating 
such that
$(k -k')v$ term can be neglected.
Hence 
the atomic polarization
should be averaged over the entire velocity
distribution such that
 \begin{equation}
  \chi
     =
        \int d(kv)~f(kv) ~\eta
        \left\{ {\rho_{ab}(kv) \over \alpha} \right\},
  \label{j-1}
 \end{equation}
where
$f(kv)$ is the velocity distribution function,
and again $\eta$ is given by
Eq.~(\ref{chi-1}).
We now consider the case where
the inhomogeneous line
is bigger than any other quantities involved
such that $W_D \gg \Omega , \gamma \gg \gamma_{bc}$,
and the condition 
$\Omega^2 \gg \gamma_{bc} \gamma$
is still satisfied.

The population distribution
in Eq.~(\ref{pop}) 
is now different for
atoms with different velocities.
As we mentioned in Sec.~II,
we then need to replace 
$\Delta_0$ with $\Delta_0 + k' v \approx \Delta_0 + k v$
for the expression of $X$ in Eq.~(\ref{xx})
such that
for a resonant driving field ($\Delta_0=0$)
we have

\begarr
X &\approx& {\gamma^2 + (k v)^2 \over 2 \gamma},
\qquad
D \approx
        2 \gamma_{bc} [\gamma^2 + (k v)^2]  + \Omega^2 \gamma  .
\label{xx-2}
\endarr

\no
Hence, for the atom with its velocity $v$,
$\rho_{ab}(kv)$ can be written 
 \begin{eqnarray}
  \rho_{ab} (kv)
    &=&
        { i \alpha \over Y }
        {1 \over 2 D}
        \Big[
             \Gamma_{c b}
          (4 \gamma X \gamma_{bc} + 2 \Omega^2 \gamma)
    \nonumber \\
    && \hspace{4em}
         - \frac{\Omega^2  4 \gamma X \gamma_{bc} }
           { \gamma + \gamma_{bc}/2 + i k v }
        \Big] , 
  \label{j-4}
 \end{eqnarray}
where
$
Y= (\gamma +\gamma_{bc}/2 +i \Delta + i k v)
(\gamma_{bc}+i\Delta) + \Omega^2
$.
Here we have assumed
$k' \approx k$ and $(k-k')v$
terms can be neglected for the copropagating fields.

\subsection{Doppler average using a Lorentzian distribution}

We now need to evaluate the expression of
susceptibility
given in Eq.~(\ref{j-1}).\no
Normally, the velocity distribution is described by
a Gaussian function
given by

\begin{equation}
f (kv) = \frac{1}{\sqrt{\pi}ku}
\exp\left[{-\frac{(kv)^2}{(ku)^2} }\right],
\label{gaussian}
\end{equation}

\no
where $u=\sqrt{2 k_B T/M}$ is the most probable
speed of the atom given by
temperature $T$ and the atomic mass $M$.
Then, the full width at half maximum is given
as $2 W_D = 2 \sqrt{\ln{2}} k u$.
However, in our analysis,
for the sake of simple
analytic expressions,
we adopt a Lorentzian distribution
of FWHM of $2W_D$,
instead of a Gaussian distribution,
such that

 \begin{equation}
  f(kv)
    =
        { W_D/\pi \over W_D^2 + (kv)^2}.
  \label{h7}
 \end{equation}

The two distributions are 
shown in Fig.~2, and there
we see that a Gaussian distribution
with the same width (FWHM $2W_D$) has
its maximum
larger than that of Lorentzian distribution
by a factor of $\sqrt{\pi \ln{2}}$.
Hence, if we multiply the factor $\sqrt{\pi \ln{2}}$
in  Eq.~(\ref{h7}), the central distribution becomes very
similar to that of Gaussian as illustrated in Fig.~2(c).

\begin{figure}[htb]
\epsfysize=5cm
\centerline{\epsfbox{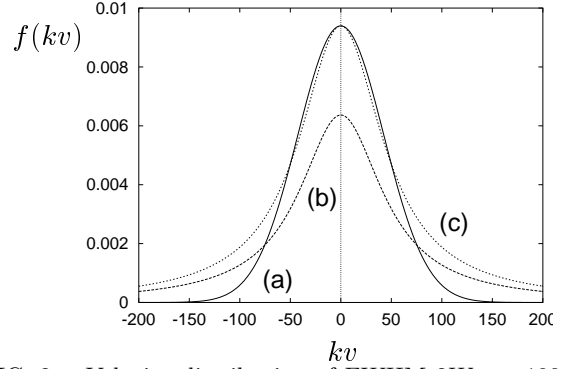}}
\caption{
\label{fig-2}
Velocity distribution of FWHM $2W_D=100 \gamma$
as a function a $kv$ in unit of $\gamma$, with
(a) a Gaussian profile of Eq.(\ref{gaussian}),
(b) a Lorentzian profile of Eq.(\ref{h7}),
(c) the plot ofr Eq.(\ref{h7})
multiplied by a factor $\sqrt{\pi \ln{2}}$.
}
\end{figure}

\v

\no
In Fig.~3 the absorption profiles are
described numerically by using the two different
distributions.  We note that
the two distributions give an almost identical
result when the factor $\sqrt{\pi \ln{2}}$
is taken into account, see Fig.~3(c).

\begin{figure}[htb]
\epsfysize=5cm
\centerline{\epsfbox{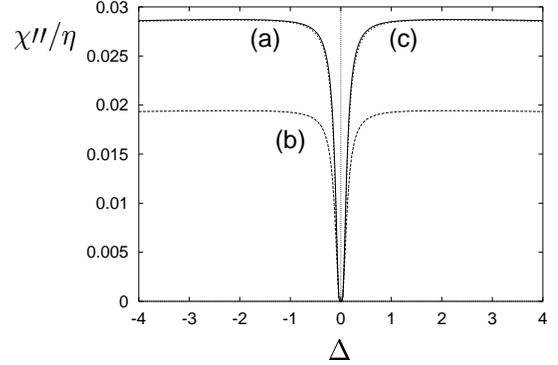}}
\caption{
\label{fig-3}
Absorption profiles ($\chi\prime\prime/\eta$)
as a function of probe field detuning 
($\Delta$ in unit of $\gamma$)
for $ 2W_D=100\gamma$, $\gamma_{bc}=10^{-3}\gamma$,
and $\Omega=2\gamma$ using
(a), (b), (c) of Fig.~2, respectively.
}
\end{figure}

Now using the distribution of Eq.~(\ref{h7}),
Eq.~(\ref{j-1})
may be considered as a contour integration
in the complex plane.
We find
three poles in the the upper half plane
at

 \begin{eqnarray}
  kv
    &=&
        \frac
         {\Delta ( \Omega^2 -\gamma_{bc}^2 - \Delta^2 )
          +i (\Delta^2 \gamma + \gamma_{bc}^2 \gamma + \gamma_{bc} \Omega^2)
         }
          {\gamma_{bc}^2 + \Delta^2}
        \;,
    \nonumber \\
    && 
         \quad
        i W_D,
    \qquad \qquad
        i\sqrt{\Omega^2 \gamma \over 2 \gamma_{bc}},
  \label{h8}
 \end{eqnarray}
and two poles in the lower half plane as

\begin{equation}
kv = - i W_D,
\qquad
 -i \sqrt{\Omega^2 \gamma \over 2 \gamma_{bc}}.
\label{h9}
\end{equation}
\no
We can see that one pole is from the
expression $Y$,
two poles ($\pm i \sqrt{\Omega^2 \gamma /2 \gamma_{bc}}$) 
are from the expression $D$ 
in Eq.~(\ref{j-4}),
and
two poles ($\pm i W_D$) are
from velocity distribution function
\cite{note-pole}.
Let us take the contour
in the lower half plane and
denote

\begin{equation}
\chi =\chi_1 + \chi_2,
\label{h10}
\end{equation}
where
$\chi_i$'s are the contributions
from the two poles
at $ - i W_D$
and
$ -i \sqrt{\Omega^2 \gamma / 2 \gamma_{bc}}$,
respectively.
For the pole at $kv =-i W_D$, we obtain

\begin{eqnarray}
\chi_1 &=&
{-i \eta \over 2 Z_1 A}
\bigg[ (B_1 - \Delta^2)
-i \Delta W_D \bigg]~
\bigg[ C_1 - i \Delta D_1 \bigg],
\label{h11}
\end{eqnarray}
where
$A$ is given by

\begarr
  A
   &= & - 2 \gamma_{bc} W_D^2  + \Omega^2 \gamma,
\label{a-value}
\endarr

\no
and

\begin{eqnarray}
Z_1 &=& (\gamma_{bc} W_D + \Omega^2 - \Delta^2)^2
+\Delta^2 W_D^2, \non \\
B_1 &=& \gamma_{bc} W_D + \Omega^2, 
\nonumber \\
C_1 &=& 2 \gamma_{bc} W_D (\gamma_{bc} W_D + \Omega^2)
-2 \gamma_{bc} \Omega^2 \gamma,
\non \\
D_1 &=& -2 \gamma_{bc} W_D^2 + 2 \Omega^2 \gamma.
\label{h13}
\end{eqnarray}
\no
For the pole at $kv = -i \sqrt{\Omega^2 \gamma / 2 \gamma_{bc}}$,
we have

\begin{eqnarray}
\chi_2 &=&
{i \eta \Omega^2 \gamma W_D \over 2 Z_2 A y}
\bigg[ (B_2 - \Delta^2)
-i \Delta y \bigg]~
\bigg[ C_2 - i \Delta \bigg],
\label{h14}
\end{eqnarray}
where
$
y =  \sqrt{\Omega^2 \gamma / 2 \gamma_{bc}}
$,
and
\begin{eqnarray}
Z_2 &=& (\gamma_{bc} y + \Omega^2 - \Delta^2)^2
+\Delta^2 y^2, 
\non \\
B_2 &=& \gamma_{bc} y + \Omega^2, 
\non \\
C_2 &=& - \gamma_{bc} + \Omega^2/y .
\label{h16}
\end{eqnarray}

\no
Note that we have assumed $\Omega^2 \gg \gamma_{bc} \gamma$,
$W_D \gg \Omega , \gamma \gg \gamma_{bc}$.

\subsection{Absorption
and dispersion at EIT resonance}

The absorption profile is now obtained
by the imaginary parts of Eqs.(\ref{h11}, \ref{h14})
as

\begin{eqnarray}
\chi_1^{\prime \prime} &=&
{-\eta \over 2 Z_1 A}
\bigg[ (B_1-\Delta^2) C_1 - \Delta^2 W_D D_1\bigg] , \nonumber \\
\chi_2^{\prime \prime} &=&
{\eta \Omega^2 \gamma W_D \over 2 Z_2 A y}
\bigg[ (B_2-\Delta^2) C_2 - \Delta^2 y \bigg].
\label{h17}
\end{eqnarray}
\no
Taking $\Delta=0$,
we found

\begin{eqnarray}
\chi_1^{\prime \prime} (\Delta=0) &=&
{- \eta \over  A }
\left[
\gamma_{bc} W_D - { \gamma_{bc} \Omega^2 \gamma \over
 + \gamma_{bc} W_D + \Omega^2}
\right],
\nonumber \\
\chi_2^{\prime \prime} (\Delta=0) &=&
{\eta \gamma_{bc} W_D \over  A }
\left[
1
- {2 \gamma_{bc} y 
\over \gamma_{bc} y + \Omega^2}
\right],
\label{h18}
\end{eqnarray}

\no
which gives the minimum value of absorption at the EIT
line center as

 \begin{equation}
  \chi^{\prime \prime} (\Delta=0)
     =
        {\eta \gamma_{bc} \over \gamma_{bc} W_D + \Omega^2}
        \left[
          {\sqrt{x} \over 1 + \sqrt{x}}
        \right] ,
   \label{min}
 \end{equation}
\no
where 

\bege
x = {\Omega^2 \gamma \over 2 \gamma_{bc} W_D^2}.
\label{parameter}
\ende

\no
We note that when $x \ll 1$,

\begin{eqnarray}
    \left. \chi^{\prime \prime}\right|_{\Delta = 0}
    &\Longrightarrow&~
        {\eta \gamma_{bc} \over \gamma_{bc} W_D + \Omega^2}
          \sqrt{x}
         ~<~   {\eta \sqrt{x} \over  W_D}
         ~\ll~
             {\eta \over W_D} ,
   \label{ab-sm} 
\end{eqnarray}

\no
and when $x \gg 1$,

\begin{eqnarray}
    \left. \chi^{\prime \prime}\right|_{\Delta = 0}
    &\Longrightarrow&~
        {\eta \gamma_{bc} \over \gamma_{bc} W_D + \Omega^2}
        ~<~
          {\eta \gamma_{bc} \over \Omega^2}
        ~\ll~ 
        {\eta \gamma \over W_D^2 } .
   \label{ab-big}
 \end{eqnarray}

\no
In both cases 
the EIT can be achieved, i.e.,
$\left. \chi^{\prime \prime}\right|_{\Delta = 0} \ll \eta /W_D$.
Therefore,
the condition for EIT is
still $\Omega^2 \gg \gamma \gamma_{bc}$,
the same as in the homogeneously broadened system.


One interesting quantity here is
the slope of the real part of the susceptibility, which
is important in precision magnetometry, and
also governs the group velocity of the probe light.
From Eqs.~(\ref{h11},\ref{h14}) the real part of the susceptibility
is found as

\begin{eqnarray}
\chi_1^{\prime} &=&
{-\eta \Delta \over 2 Z_1 A}
\bigg[ W_D C_1 + D_1(B_1 - \Delta^2) \bigg], \nonumber \\
\chi_2^{\prime} &=&
{\eta \Omega^2 \gamma W_D \over 2 Z_2 A y}
\Delta
\bigg[ C_2  y + (B_2 - \Delta^2) \bigg],
\label{h20}
\end{eqnarray}
and its derivative at resonance
is given by
\begin{equation}
\left. \frac{\partial \chi_1^{\prime}}
{\partial \Delta}\right|_{\Delta = 0}
= -{\eta \gamma \over A},
\qquad
\left. \frac{\partial \chi_2^{\prime}}
{\partial \Delta}\right|_{\Delta = 0}
= {\eta \sqrt{2 \gamma_{bc} \gamma W_D^2/ \Omega^2} \over A}.
\label{h21}
\end{equation}

\no
Hence,
we obtained the slope of $\chi^{\prime}$ at $\Delta = 0$
as
\begin{eqnarray}
\left.
\frac
{d \chi^{\prime}}
{d \Delta}\right|_{\Delta = 0}
&=& - {\eta \over \Omega^2}~ 
{ \sqrt{x} \over 1 + \sqrt{x} }.
\label{j-11}
\end{eqnarray}

\no
Therefore, when $x \gg 1$,
it approaches to ${\eta /\Omega^2}$
and when $x \ll 1$,
it goes as $(\eta /\Omega^2) \sqrt{x}$.
We note that, 
under the EIT condition $\Omega^2 \gg \gamma_{bc} \gamma$,
$(\eta /\Omega^2) \sqrt{x}$ is
still much larger than 
$(\eta /\Omega^2) (\gamma/W_D)$.

\begin{eqnarray}
\left.
\frac
{d \chi^{\prime}}
{d \Delta}\right|_{\Delta = 0}
    &\Longrightarrow&~
 - {\eta \over \Omega }~ 
{ 1 \over \sqrt{2 \gamma_{bc} \gamma} } 
\left( {\gamma \over W_D}\right).
\label{dis-sm}
\end{eqnarray}

\subsection{Transmission width of EIT resonance}

In order to estimate the linewidth of EIT
we take the same procedure
as in Sec.~II:
First, we find that the maximum of $\chi^{\prime \prime}$
as $\chi_{\rm max} \approx \eta/W_D$ at $\Delta \approx \pm \Omega$.
Then, we evaluate $\Delta$ which defines
$\Gamma_{EIT}$ as
\bege
\chi^{\prime \prime} (\Delta=\Gamma_{EIT}) = \eta/2W_D.
\ende
\no
By Eq.~(\ref{h17}) it readily gives the
following equation:

\begin{equation}
\Delta^4 - {2 \gamma_{bc} \Omega^2 \over \gamma }
{2 \gamma_{bc} W_D^2 + \Omega^2 \gamma \over
2 \gamma_{bc} W_D^2 } \Delta^2
-
{2 \gamma_{bc} \Omega^2 \over \gamma }
{\Omega^4 \ W_D^2}
= 0,
\label{h19}
\end{equation}
which yields 
the half width of the EIT
for the Doppler broadened
system given by

\begin{eqnarray}
\Gamma_{EIT}^2
&=& {\gamma_{bc} \over \gamma} ~\Omega^2
(1+x)\left[
1+ \left\{ 1+ {4x \over (1+x)^2 }\right\}^{1/2}
\right] ,
\non \\
&\approx& 
{2 \gamma_{bc} \over \gamma} ~\Omega^2 (1 +x) ,
\label{j-9}
\end{eqnarray}
where
$
x = {\Omega^2 \gamma / 2 \gamma_{bc} W_D^2}
$ given by Eq.~(\ref{parameter}).
Now 
if we define 
a saturation intensity as
 
\bege
\Omega_{s}^2
= {2 \gamma_{bc} W_D^2 \over \gamma},
\ende
the linewidth expression can be written as

\begin{eqnarray}
\Gamma_{EIT}
&\approx& 
\left[
\sqrt{2 \gamma_{bc} \over \gamma} ~\Omega 
\right]
\sqrt{ 1 + {\Omega^2 \over \Omega_s^2} }
\label{fj-limit}
\end{eqnarray}

\no
Here we can see that
in the limit $\Omega \ll \Omega_s$
$\Gamma_{EIT}$ is proportional
to the Rabi frequency of the driving field.
Such a linewidth was predicted
by Feld and Javan in the study of
laser induced line narrowing \cite{feld69}.
On the other hand,
in the limit $\Omega \gg \Omega_s$
$\Gamma_{EIT}$ is proportional to
the intensity of the driving field
($\Omega^2/W_D$).
This power broadening feature is
shown in Fig.~4.

\begin{figure}[htb]
\epsfysize=5cm
\centerline{\epsfbox{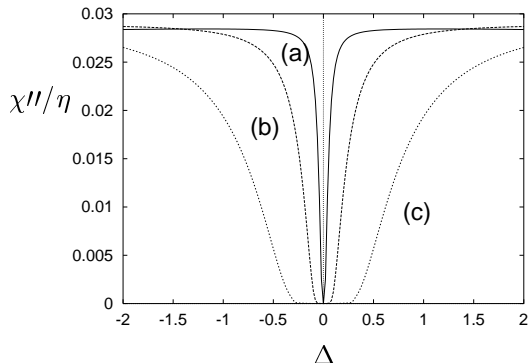}}
\caption{
\label{fig-4}
Absorption profile for 
$2W_D=100\gamma$,
$\gamma_{bc} = 10^{-3} \gamma$.
(a) $\Omega = \gamma$,
(b) $\Omega = 3\gamma$,
(c) $\Omega = 6\gamma$.
Note that $\sqrt{\gamma_{bc}\gamma} \sim 0.03 \gamma$
and $\Omega_s \sim 4.5 \gamma$.
}
\end{figure}

The expression of Eq.~(\ref{fj-limit})
shows a reminiscence
of power broadening factor
in the description of hole burning\cite{yariv89}.
In place of the homogeneous linewidth
in the expression of hole burning,
here we have an effective width
which is determined by the spectral
packet involved in population trapping \cite{javan00}.

\subsection{The role of optical pumping}

We have seen that the parameter $x=\Omega/\Omega_s$
plays an important role in the case of inhomogeneously
broadened medium.
Let us here examine 
the physical meaning of the parameter.

Suppose the system is
homogeneously broadened.
When the driving is on resonance,
the optical pumping rate from the level $c$
is then order of $\Omega^2 / \gamma$, 
as given in Eq.~(\ref{op}).
A complete optical
pumping within the homogeneous linewidth,
is then possible if 
this rate is larger than the pumping
from level $b$ to $c$:
$\Omega^2 / \gamma \gg \gamma_{bc}$.
This, in turn,
gives the EIT condition. 
When we have the driving field detuned by $\Delta_0$,
the optical pumping rate
decreases by a factor of $\gamma^2/(\gamma^2 + \Delta_0^2)$.
Again for a complete optical pumping 
we need $\Omega^2 \gamma / [\gamma^2 + \Delta_0^2] \gg \gamma_{bc}$.

If we now assume that 
we have the resonant driving 
field $\Delta_0 =0$ again,
and, instead,
the atoms are moving.
Then, for atoms with velocity $v$,
the optical pumping rate becomes
$\Omega^2 \gamma / [\gamma^2 + (kv)^2]$.
Then, on the average,
to have a complete optical pumping
in a Doppler broadened system
we need to require
$\Omega^2 \gamma / (\gamma^2 + W_D^2) \gg \gamma_{bc}$,
which corresponds to $x \gg 1$  (assuming $W_D \gg \gamma$), 
i.e. $\Omega \gg \Omega_s \equiv  2\gamma_{bc}W_D^2/\gamma$.
Hence, the parameter $x= \Omega^2/\Omega_s^2$ represents
the degree of saturation in $b \leftrightarrow c$
transition, or
the degree of optical pumping from the level $c$
to $b$ within the inhomogeneous linewidth.

\section{Comparison with an open system description}

In this section we examine the case of an open system
and show that the result is essentially
the same as our model of a closed system.
The open system is modeled for the atoms that
are coming in
and out of the interaction (with the radiation fields)
region.
Although in such a case all the levels have the same decay rate
(say, $\gamma_{bc}$),
the upper level can decay much faster than
the time of flight through the interaction region
(for example, radiative decay or collisional decay).
Hence we assume that
the lower levels $b$ and $c$ decay with rate
$\gamma_{bc}$ and the upper level $a$ decays
with rate $\gamma_a$ which is much bigger than
$\gamma_{bc}$ (see Fig.~5).

\begin{figure}[htb]
\epsfysize=4cm
\centerline{\epsfbox{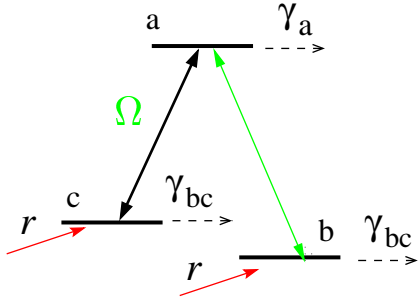}}
\caption{
\label{fig-5}
Model scheme of the open system.
Upper level $a$ decays with rate $\gamma_a$
Lower levels $b$ and $c$
decay with the same rate $\gamma_{bc}$.
Atoms are pumped at a rate 
$r$ equally to the lower levels.
}
\end{figure}

Furthermore, for simplicity, we 
assume that the atoms are coming
into the interaction region with a same rate for
the lower levels.
Under these assumption, the equation of motion for the density
matrix elements can be written as

\begin{mathletters}
\label{a1}
\begarr
{\dot\rho}_{ab}&=& -  {\Gamma}_{ab}\rho_{ab}
           -i \alpha (\rho_{aa}-\rho_{bb}) +i \Omega \rho_{cb},
\label{a} \\
{\dot\rho}_{cb}&=&  -  {\Gamma}_{cb}\rho_{cb}
           -i \alpha \rho_{ca} +i \Omega \rho_{ab},
\label{b} \\
{\dot\rho}_{ac}&=&  -  {\Gamma}_{ac}\rho_{ac}
           -i \alpha \rho_{bc} -i \Omega (\rho_{aa}-\rho_{cc}),
\label{c} \\
{\dot\rho}_{aa}&=& -  \gamma_a \rho_{aa}
           - i\alpha( \rho_{ab} -\rho_{ba} ) 
           - i\Omega (\rho_{ac}- \rho_{ca}) ,
\label{d} \\
{\dot\rho}_{bb}&=&  r -  \gamma_{bc} \rho_{bb}
           - i\alpha( \rho_{ab} -\rho_{ba} ) 
           - i\Omega (\rho_{ac}- \rho_{ca}) ,
\label{e}\\
{\dot\rho}_{cc}&=&  r -\gamma_{bc} \rho_{cc}
           - i\Omega (\rho_{ca}- \rho_{ac}) .
\label{f}
\endarr
\end{mathletters}

\no
Here the notations are the same as 
Eq.~(\ref{h2}).
Note that now we have
$
\Gamma_{ac} = (\gamma_a + \gamma_{bc}) / 2 + i \Delta_0
$,
which gives

\bege
  X' \equiv
  {
   [(\gamma_a/2 + \gamma_{bc}/2)^2 + \Delta_0^2] \over
   2 (\gamma_a/2 +\gamma_{bc}/2) }.
\label{xprime}
\ende

\no
Again
if we assume that $\gamma_a \gg \gamma_{bc}$,
the populations are found
as

\begin{eqnarray}
\rho_{aa}^{(0)} -\rho_{cc}^{(0)} 
 &\approx& - {  \gamma_{bc} \gamma_a X'  \over 2D' },
\quad
\rho_{bb}^{(0)} 
\approx { \gamma_{bc} \gamma_a X'  + \Omega^2 \gamma_a \over 2D' } ,
\label{a6}
\end{eqnarray}

\no
where

\begarr
X' &\approx& {(\gamma_a/2)^2 + \Delta_0^2 \over \gamma_a},
\qquad
D' \approx
        \gamma_{bc} \gamma_a X' + \Omega^2 \gamma_a  .
\label{a7}
\endarr

\no
Comparing Eq.~(\ref{a6}) with (\ref{pop}),
we can see that the population distribution is
almost identical to the one for the model of closed
system.

Furthermore,
the expression for $\rho_{ab}^{(1)}$
is identical to the one for the closed system
given in Eq.~(\ref{h6}). 
Let us then recall Eq.~(\ref{a-value}) saying that
$A = - 2 \gamma_{bc} W_D^2  + \Omega^2 \gamma$,
which is obtained by putting $-i W_D$ to $\Delta_0$
in the expression of $D$ in Eq.~(\ref{h4}).
The sign of $A$ determines whether
the crucial parameter $x$ is $> 1$ or $ <1$.
Similarly, here for the open system,
when we put $\Delta_0=-iW_D$, we can define $A'$
as $A' \approx -\gamma_{bc} W_D^2 + \Omega^2 \gamma_a$
such that we have
$x' = \Omega \gamma_a / \gamma_{bc} W_D^2$
as the parameter which plays the same role as $x$
in Eq.~(\ref{xx}).
Hence, by replacing $\gamma_a \Rightarrow 2 \gamma$,
we have the open system description
almost identical to the description for
our model scheme of the closed system
A detailed analysis of the open system
will be presented elsewhere \cite{rost00}.

\section{Summary}

In this paper, we have studied
the transmission width of EIT
in a three-level $\Lambda$ system.
The Doppler averaged susceptibility
is found by
using a Lorentzian velocity distribution
rather than the Gaussian distribution.
Then we have shown
the requirement for achieving EIT, 
and 
the analytic expression of the EIT linewidth.
The saturation intensity $\Omega_s^2$
defines the degree of optical pumping as $\Omega^2/\Omega_s^2$,
and represents
the {\it condition} under which the broadening is either
linear or quadratic in the Rabi frequency of the driving field.

\section*{Acknowledgments}

We would like to thank B.\ G.\ Englert,
O.\ Kocharovskaya, A.\ B.\ Matsko,
I.~Protsenko, M.~O.~Scully, 
V.~L.~Velichansky, and A.~S.~Zibrov for helpful discussions.
This work was supported by the Office of Naval Research,
the National Science Foundation,
and the Welch Foundation.

\end{multicols}

\begin{thebibliography}{99}
\bibitem{note-pole}
Note that the expression 
$
 - \Omega^2  4 \gamma X \gamma_{bc}/ (\gamma + \gamma_{bc}/2 + i k v )
$
in Eq.~(\ref{j-4})
does not have a pole
as we recall the original form of $X$
in Eq.~(\ref{x}).
\end{thebibliography}
\end{document}